\documentclass{aastex}
\usepackage{spr-astr-addons}
\usepackage{url}

\RequirePackage{color}

\begin{document}

\title{Star Cluster Formation and Star Formation: \\
The Role of Environment and Star Formation Efficiencies}
%\slugcomment{Not to appear in Nonlearned J., 45.}
%% Running heads
\shorttitle{Star and Star Cluster Formation}
\shortauthors{Uta Fritze}

\author{Uta Fritze}
\affil{University of Hertfordshire}
%\email{\emaila}

%\altaffiltext{1}{First Alternate Affilation.}
%\altaffiltext{2}{Second Alternate Affilation.}
%\altaffiltext{3}{Third Alternate Affilation.}

\begin{abstract}
Analyzing global starburst properties in various kinds of starburst and post-starburst galaxies and relating them to the properties of the star cluster populations they form, I explore the conditions for the formation of massive, compact, long-lived star clusters. 
The aim is to find out whether the relative amount of star formation that goes into star cluster formation as opposed to field star formation, and into the formation of massive long-lived clusters in particular, is universal or scales with star formation rate, burst strength, star formation efficiency, galaxy or gas mass, and whether or not there are special conditions or some threshold for the formation of star clusters that merit to be called globular clusters a few gigayears later. 
\end{abstract}

\keywords{star formation; star cluster formation; star formation efficiencies; environment}

%\section*{}
%\label{sec:intro}

\section{Motivation}

Star Cluster (SC) formation is a major or even dominant mode of all star formation (SF) and occurs in very different environments. This immediately raises the question whether young star clusters (YSCs) forming in different environments are similar or different. 

SCs are not only interesting in their own right, but bear considerable power as tracers of SF in their parent galaxies. YSCs trace the spatial distribution of SF and its recent history within a galaxy, old globular clusters (GCs) trace violent SF phases in their parent galaxy -- all the way back to the very onset of SF, i.e. over a Hubble time. But SCs also fade and dissolve. In actively star forming galaxies, the youngest SCs may still be embedded in their natal dust clouds while part of the older SCs are already gone, dissolved and/or faded below detection. It is therefore important to take these processes into account when comparing SC populations in different galaxies. They depend on the initial properties of individual SCs, their masses, radii, abundances, stellar IMF, and of the SC population, i.e. the luminosity function, the mass function, distribution of radii, ages, etc. 

That SC formation is an important mode of SF in starbursts was shown on the example of the Tadpole and Mice interacting galaxies. A pixel-by-pixel analysis of ACS data ($BVRI$) with GALEV evolutionary synthesis models showed that $\sim 70$ \% of the blue light is emitted by YSCs as opposed to only $\sim 30$ \% coming from field stars. We estimated that more than 35 \% of all SF went into the formation of YSCs, not only in the main bodies of these two galaxies, but all along their extended tidal tails \citep{deGrijs+03a}. Clearly, this analysis needs to be extended to different types of galaxies, starburst/non-starburst, dwarf/normal, gas-rich/gas-poor, interacting/non-interacting, in various stages of the interaction, etc. to explore the systematics. There are indications from \cite*{Meurer+95}'s work that the contribution from YSCs to the UV light of a galaxy increases with increasing UV surface brightness, which itself is a measure of SF intensity. A burning question that we are currently exploring is whether the amount of SF that goes into massive long-lived SCs in relation to the amount of SF that goes into low-mass short-lived SCs and field stars, increases with increasing overall SFR, with bursts strength, or with star formation efficiency (SFE). If so, the immediate next question is whether this is a continuous increase or rather a threshold effect in the sense that only above a certain SF intensity or efficiency the massive long-lived SCs are formed that later are called GCs. 

SCs form in very different environments, in normal star-forming galaxies, as e.g. M51 or NGC 5236 (cf. \cite*{Larsen04}, \cite*{Mora+07}, in dwarf galaxy starbursts like NGC 1569 \citep{Anders+04b}, in interacting gas-rich galaxies like NGC 4038/39 (the Antennae) \citep{Whitmore+95,Whitmore+05,Anders+07}. Slightly older and intermediate-age SCs are observed in post-starburst merger remnants like NGC 7252 \citep{Whitmore+93,FB95,Miller+97,SchweizerSeitzer98} and dynamically young ellipticals like NGC 1316 \citep{Goudfrooij+01b,Goudfrooij+04,Goudfrooij+07}, respectively. They can form all over the main body of a galaxy, as e.g. in the Antennae or NGC 7252, in and around a starburst nucleus, as e.g. in Arp 220 or NGC 6240 \citep{Shioya+01,Pasquali+03}, all along some -- but not all -- extended tidal features \citep{Knierman+03,deGrijs+03a,Trancho+07a}, as well as in group environments like Stephan's quintett \citep{Gallagher+01}. These environments cover a huge range in terms of density, kinetic temperature, chemical abundances and it is by no means obvious whether or not all these SCs are similar or different, individually or as a population. Related questions are where, when, and how GC are formed and what a young GC looks like. Or how to tell apart YSCs into long-lived and short-lived ones -- by mass, concentration, mass function, ...? 

Current cluster formation models require exceptionally high SF efficiencies ${\rm SFE~:=~M_{\ast}/M_{gas}~>30~\%}$ as a prerequisite for the formation of massive strongly bound and long-term stable SCs, i.e. for the formation of young GCs \citep{Brown+95,Burkert+96,ElmegreenEfremov97,Li+04}. On a global scale, SF efficiencies in normal spiral and irregular galaxies, as well as in starbursting dwarf galaxies are of order $0.1 - 3 \% $ \citep{Krueger+95}. On the smaller scale of molecular clouds in the Milky Way, the SF efficiency is of the same order of magnitude and so is the mass ratio between the molecular cloud core and the entire molecular cloud. No GC formation is therefore expected in spirals, irregulars or star-bursting dwarf galaxies by today. In giant gas-rich interacting galaxies, on the other hand, SF efficiencies of order $10 - 50 \%$ are reported on global scales, and of order $30 - 90 \%$ on nuclear scales of a few hundred pc up to $\sim 1$ kpc. In those systems, GC formation should be possible. The fact that different submm lines (CO(1-0), HCN(1-0), CS(1-0)) trace molecular gas at different densities (${\rm n \geq 100, n~\geq 3 \cdot 10^4, n~\sim 10^5~cm^{-3}}$), has allowed to see that while in the Milky Way and other nearby galaxies only a small fraction ($0.1 - 3 \%$) of the mass of a molecular cloud makes up its high density core, the situation is drastically different in Ultraluminous Infrared galaxies (ULIRGs), which all are late stages of massive gas-rich mergers with strong nuclear (few 100 pc) starbursts. In those ULIRGs, almost all the molecular gas in the central starburst region is at the high densities of molecular cloud {\em cores}, indicating that the molecular cloud structure must be very different from what we know in our Galaxy. The entire nuclear region is just one supergiant molecular cloud core, seriously raising the question whether the star and SC formation processes can be the same as in normal galaxies, not to mention the situation in extended, expanding, low-density tidal structures in the outskirts of other interacting galaxies. 

In any case, before ALMA becomes operational, the YSCs forming in these different types of environments are our best proxy to the molecular cloud structure. In the Milky Way, molecular cloud cores, molecular clouds, and YSCs all feature power-law mass functions, suggesting scale-free self-similar evolution. Not even for the closest massive merger, the Antennae, can we presently determine the molecular cloud or cloud core mass functions (cf. \cite*{Wilson+03}). The masses of YSCs and the shape of their mass function is all we can access \citep{Wilson+06,Anders+07}. \cite*{GaoSolomon04} and \cite*{Solomon+92} have shown that for all galaxies -- from Blue Compact Dwarfs to spirals and ULIRGs -- there is a tight correlation between SFR, as derived from far-infrared luminosity, and the mass in molecular cloud cores, as derived from the HCN luminosity. They also find the SF efficiency to be proportional to the mass ratio of molecular gas at core and normal densities, i.e. to the ratio between HCN or CS luminosity and CO luminosity. The highest density molecular gas in all these environments is transformed into stars with almost 100 \% efficiency and it is the amount of gas at those high densities that controls SF. The fraction of molecular gas at the highest densities therefore defines the SF efficiency. 

The high ambient pressure building up in the course of massive gas-rich mergers can drive up SF efficiencies by $1-2$ orders of magnitude by compressing molecular clouds, increasing their masses and, in particular, their core mass fractions. \cite*{JogDas92,JogDas96} have shown that the ISM pressure during mergers can easily become $3-4$ times higher than the typical internal molecular cloud pressure, raising the SF efficiency to $70-90 \%$. This leads us to expect that the relative amount of SF that goes into the formation of massive, strongly bound young GCs in relation to the amount of SF that goes into field stars and low-mass, short-lived clusters is enhanced in massive gas-rich mergers. 

\section{Results so far ...}
\subsection{Analysis method}
Before I turn to the results obtained so far for SCs and SC populations in different environments, I briefly recall our GALEV evolutionary synthesis models and the dedicated analysis tools we use in our analysis of SC systems. GALEV models in the first place describe the spectral evolution of SCs of various metallicities ${\rm -1.7 \leq [Fe/H] \leq +0.4}$ over the age range from 4 Myr through 13 Gyr, including gaseous emission, which significantly affects broad band luminosities and colours during early evolutionary stages (see \cite*{AndersFritze03} for details). Spectra are then folded with filter functions for any desired filter system to yield the photometric evolution. This is important in order to avoid uncertainties from {\em a posteriori} transformations between filter systems. Models well reproduce empirical colour -- metallicity calibrations over their range of validity and indicate significant deviations from their linear behaviour towards higher metallicties. We showed that transformations from colour to metallicity are significantly age-dependent and that transformations from colour to age are significantly metallicity-dependent \citep{Schulz+02}. The effect of dust absorption is included in GALEV models assuming a starburst extinction law \citep{Calzetti+00} for a range of values for $E(B-V)$ ($0 \leq E(B-V) \leq 1~$mag). GALEV models also include the full set of Lick spectral absorption indices on the basis of empirical calibrations for the indices in terms of stellar parameters for every individual cluster star ${\rm T_{eff},~log~g,~[Fe/H]}$ as given by \citep{Gorgas+93} and \citep{Worthey+94}. We showed that the transformation from the age-sensitive Lick index H$_{\beta}$ to age is significantly metallicity-dependent and that the transformation from the metallicity-sensitive Lick indices (Mgb, Mg$_2$, [MgFe], $\dots$) to metallicity is age-dependent for ages $\leq 10$ Gyr \citep{Kurth+99,LillyFritze06}. 

Our analysis methods use the full information from multi-band imaging ($UV,~U,~B,\dots,~NIR$) or/and Lick spectroscopy available for a SC system, compare them to a large grid of over 100.000 GALEV models in terms of Spectral Energy Distributions (SEDs), Lick indices, or a combination of both (cf. \cite*{Anders+04a}, \cite*{LillyFritze06}, Lilly \& Fritze 2008, {\em submitted}). SEDs, we recall, are sets of magnitudes in a number of filters from short to long wavelengths, e.g. $U ~ \dots ~ K$. Our  analysis tools not only determine {\em the} best fit model but attribute probabilities to all models that allow us to determine the $1 \sigma$ uncertainties for all the SC parameters they return: age, metallicity, $E(B-V)$, and mass. Extensive tests with artificial SCs have shown that $UV$ or $U-$band observations are essential for age dating of YSCs and a NIR band is important to obtain accurate metallicities. For YSCs in dusty galaxies four passbands including $UV/U$ and $H$ or $K$ with observational uncertainties $\leq 0.05$ mag in the UV/optical and $\leq 0.1$ mag in the NIR allow to largely disentangle ages and and metallicities and to obtain ages to $\Delta~{\rm age / age} \leq 0.3$ and metallicties to $\pm 0.2$ dex. For intermediate-age SCs or old GCs in dust-free environments, three passbands, again ranging from $U$ or $B$ through $H$ or $K$ are enough \citep{Anders+04a,deGrijs+03c}. 

\subsection{Star cluster formation in dwarf galaxy starbursts}

Applying our SED analysis tool to HST WFPC2 and NICMOS archival data for some 170 compact YSCs that we identified in the not apparently interacting dwarf starburst galaxy NGC 1569, we obtained masses for the bulk of its YSCs in the range ${\rm 10^3 - 10^4~M_{\odot}}$. Only a handful of these, including the 3 previously known so-called Super Star Clusters, have masses above a few ${\rm 10^5~M_{\odot}}$, i.e. in the range of GC masses. We conclude that this strongly starbursting but not apparently interacting dwarf galaxy does not form any new GCs, or, at most, very few \citep{Anders+04b}. 

\subsection{Star cluster formation in the merger remnant NGC 7252}

For the starburst in the massive gas-rich spiral -- spiral merger remnant NGC 7252, we could estimate the SF efficiency very conservatively to be at least 35 \%. This estimate was based on the amount of new stars formed during the burst, as obtained from the deep Balmer absorption lines in the overall spectrum, and a very generous estimate of the gas mass available in the two Sc-type progenitor spirals, of which the ample HI still observed all along the extended tidal tails is the proof \citep{FG94a,FG94b}. Such a high SF efficiency should allow for the formation of massive, compact, strongly bound GCs. HST observations indeed revealed a rich population of compact SCs with ages in the range $600-900$ Myr and metallicities close to solar \citep{FB95}. They apparently have survived many internal crossing times and the most critical phase in their lives, the infant mortality phase after expulsion of the gas left over at their formation when the first SNe went off, and they are still compact and bound. This is particularly impressive since all this happened during the violent relaxation phase that restructured their parent galaxy from two spiral disks into a spherical configuration featuring a de Vaucouleurs profile \citep{Schweizer06}. These young GCs have all chances to survive another Hubble time. They have masses in the range ${\rm 10^5 - 10^6~M_{\odot}}$ with cluster W3 even reaching ${\rm 7-8~M_{\odot}}$ \citep{FB95,Maraston+04}. Enough of those young GCs survived until today to secure the merger remnant the typical GC specific frequency of an elliptical galaxy, which is twice as high when defined in terms of number of GCs in relation to galaxy total {\em mass} as for an average spiral \citep{ZepfAshman93}. I.e. during the strong global starburst accompanying the merger that transformed two bright Sc galaxies into a dynamically still young elliptical, a number of secondary GCs has formed that is comparable to the number of preexisting GCs in both progenitor spirals. 

\subsection{Star cluster formation in the ongoing merger NGC 4038/39}

The ongoing gas-rich spiral -- spiral merger NGC 4038/39, the Antennae, forms a rich YSC system. It is not possible to tell apart the YSCs into short-lived and long-lived ones. In any kind of observationally accessible parameter (mass, half-light radius) or parameter combination, these YSCs form a continuous distribution. In a very careful analysis of this SC system in formation, including conservative SC identification, accurate aperture corrections for SC sizes and photometry, careful completeness analysis, extensive statistical tests and likelihood evaluations by Monte Carlo simulations, we could show that the luminosity function of the SC system features a turnover with 99.5 \% significance \citep{Anders+07}. In this respect, the YSC luminosity function in this ongoing merger differs from luminosity functions of YSCs in dwarf galaxies, spirals, and isolated starbursts, which all are power laws. It also differs from the power laws found for the {\em mass} functions of molecular clouds and molecular cloud cores in undisturbed galaxies. Mass functions of molecular clouds and molecular cloud cores in gas-rich interacting galaxies cannot yet be measured, they will have to await ALMA. Unfortunately, it is not straightforward to transform the YSC luminosity function into a mass function, since it is not clear how to translate the completeness limit in luminosity into a completeness limit in mass during the rapid luminosity evolution of YSCs. The obvious way to evaluate  the mass function in small age bins suffers from low statistical significance. Repeating our accurate analysis on the ACS data covering a larger FoV with better sampling might be a promising way to go. In any case, the clear turn-over in the luminosity function could be a hint that the amount of SF that goes into massive SCs relative to the amount of SF that goes into low-mass SCs might be higher in this gas-rich merger than in other environments. It will be very interesting to check with ALMA our expectation that the molecular cloud structure and mass spectrum in this major merger are different from what they are in the Milky Way, i.e. closer to what is observed in terms of integrated light from higher and lower density molecular gas L(CS, HCN)/L(CO) in ULIRGs. NGC 4038/39 is currently a LIRG (${\rm L_{IR} > 10^{11}~L_{\odot}}$) and will probably further increase its SFR close to final merging. There is observational evidence for very large amounts of molecular gas at the level of about twice the total gas mass (HI $+~{\rm H_2}$) in the Milky Way \citep{Gao+01} and for extremely massive concentrations of it \citep{Wilson+03} with low kinetic temperature \citep{Schulz+07}. Shocked gas, on the other hand, is found displaced from the regions of high present SF, i.e. most probably due to the collision of the two galaxies \citep{Haas+05}. The exceptionally high magnetic field strength that \cite*{HummelHulst86} measured over an extended region also suggests compression of the ISM. The fraction of {\em very} dense molecular gas as seen in ${\rm L_{HCN}}$ is still low, as well as the SF efficiency, wherefrom \cite*{Gao+01} conclude that the bulk of the starburst is yet to come as the two nuclei merge, probably driving the Antennae above the ULIRG threshold in terms of IR luminosity. 

We speculate that if the turnover in the luminosity function would reflect a turnover in the underlying mass function, then this would tie in nicely with the recent result obtained by \cite*{ParmentierGilmore05,ParmentierGilmore07} that the Milky Way GC system originally must have had a mass spectrum with a turnover around ${\rm 10^5~M_{\odot}}$.  

\section{Conclusions so far ...}

{\bf We so far conclude that starbursts in non-interacting dwarf galaxies do not form substantial populations of massive, long-lived YSCs that could evolve into GC, while massive gas-rich mergers do.} 

NGC 7252 is not the only example of a merger that no doubt has produced a new generation of GCs. The $\sim 1 - 3$ Gyr old merger remnants NGC 3921 \citep{Schweizer+96}, NGC 34 \citep{SchweizerSeitzer07}, and NGC 1316 \citep{Goudfrooij+01a,Goudfrooij+01b,Goudfrooij+04,Goudfrooij+07} as well feature young GC populations formed during the mergers. The metallicities of these newly formed GCs agree well with expectations on the basis of spiral galaxy ISM properties. Evolutionary synthesis models predict these SCs to take on the optical colours of the red-peak GC widely observed in E/S0 galaxies by the time the tidal features indicative of the merger origin will have vanished. They also predict that they should readily be detectable against other populations of red GCs in combined optical and NIR observations \citep{Fritze04} (see also R. Kotulla, this volume). 
 
No example of a clearly merging gas-rich dwarf galaxy pair, nor of an accretion of a gas-rich dwarf by an elliptical or S0 has as yet been studied to check whether those would also give rise to new GC populations. 

\section{Star cluster formation in normal spirals}

When it comes to the first detailed analysis of so-called Super Star Clusters in normal actively star-forming Sbc $-$ Sd type spirals, the situation gets embarrassing. 
\cite*{Larsen04} (L04) presents ground-based and HST multi-band photometric data for a sample of 17 non-interacting actively star forming face-on spirals, in which he identifies between 7 and 149 YSCs that he calls Super Star Clusters (SSCs), and that we are currently analyzing in the way described above. All of them are compact with radii in the range of 3 to 10 pc. 

First of all, I'd like to caution the notion SSC, since it only refers to {\em luminosity} and not to {\em mass}. The term SSC goes back to van den Bergh \citep{vandenBergh71} who referred to SCs much brighter than the brightest open clusters known in the Milky Way by that time. Meanwhile, however, we know that other galaxies (e.g. the LMC) can have much richer SCs than our Milky Way and we have detailed evolutionary synthesis models that show how strongly SCs fade, in particular during their youngest stages. Depending on metallicity, a SC fades by $\sim 4 - 5$ mag in V during the first few hundred Myr -- alone through stellar evolution effects, i.e. with the stellar-dynamical mass loss not yet included. A very luminous SC therefore need not necessarily be extremely massive. Even at relatively modest masses, SCs can be {\em very bright} and look like SSCs as long as they are {\em very young}. We therefore strongly suggest to refer to {\em masses} rather than to {\em luminosities} for YSCs. 

\subsection{YSC masses}

The two brightest of L04's YSCs have spectroscopic masses of $4 \cdot 10^5$ and ${\rm 2 \cdot 10^6 ~ M_{\odot}}$ \citep{Larsen+04}. Being so few, these two  could, however, be singular outstanding SCs that formed by some local extreme compression of molecular gas, e.g. from a supergiant molecular cloud compressed between expanding shells. 

Our preliminary analysis of the 324 YSCs in those six of L04's galaxies that have $\geq 30$ YSCs each, revealed that a significant number of $\sim 70$ YSCs with ages $\geq 50$ Myr have masses ${\rm \geq 10^5~M_{\odot}}$. We have chosen a generous lower age limit of 50 Myr to be sure that the YSCs have already survived the most dangerous phase in their lives, the infant mortality phase after the first SNe have expelled the left-over gas and the subsequent dynamical rearrangement to the change in the potential (cf. Lamers, {\em this volume}). And we concentrate on YSCs with masses ${\rm \geq 10^5~M_{\odot}}$ since those have fair survival chances for the forthcoming Gyrs according to present knowledge \citep{BoutloukosLamers03} and they have masses in the range of GCs, hence merit to be called young GCs. 

\subsection{The trouble}
This result that apparently undisturbed and not currently starbursting Sbc...Sd-type spirals form SCs which have all the properties of young GCs is surprising and presents a challenge to our current understanding of SC and GC formation and evolution. \cite*{Larsen04} argues that these spirals are not currently in any particularly active state of SF, they look undisturbed and feature nice disks with regular spiral structure. The age distributions that we obtain for their YSCs support this argument. 

But how can these undisturbed spirals afford the high SF efficiencies that current theories for GC formation require? And where are the successors of previous generations of this type of SCs? Do normal actively star-forming spirals feature continuous age distributions among their GC systems? Is our Milky Way (and M31) special in this respect? Can intermediate-age GCs have escaped our detection so far? Or are these actively star-forming galaxies at the same time particularly hostile to their YSCs and destroy them beyond our current estimates? 

We discuss this in more depth in Fritze et al., {\em in prep}. 

In any case, our results require a careful reconsideration of currently accepted concepts of SC formation, evolution, and destruction.   

\acknowledgments It is a pleasure to thank the organisers for a very well-organized and inspiring conference in a particularly splendid location and to thank P. Anders, R. de Grijs, S. Larsen, and R. Kotulla for their contributions to the work presented here. 
 
\bibliographystyle{spr-mp-nameyear-cnd}
\bibliography{Fritze}

\end{document}